\documentclass{SCGE}
\usepackage[colorlinks,linkcolor=blue,citecolor=blue,urlcolor=blue]{hyperref}
\usepackage{graphicx} 
\usepackage{epstopdf}
\usepackage{multirow}
\usepackage[T1]{fontenc}
\usepackage{ulem}
\usepackage{lipsum}
\usepackage{subfigure}
\usepackage{amsmath}
\usepackage{arydshln}
\usepackage{booktabs}
\usepackage{stfloats}
\usepackage{comment}
\usepackage{threeparttable}
\usepackage{xcolor}

\let\citedash\relax
\makeatletter \providecommand{\citedash}{\hbox{-}\penalty\@m}

\begin{document}
\begin{picture}(0,0){\rm
\put(0,-20){\makebox[160truemm][l]{\bf {\sanhao\raisebox{2pt}{.}}
Article  {\sanhao\raisebox{1.5pt}{.}}}}}
\put(0,-34){\jiuwuhao {\textcolor[rgb]{0.5,0.5,0.5}{\sf 
}}}
\end{picture}
\def\bm{\boldsymbol}

\def\dl{\displaystyle}
\def\du{\end{document}}
\def\d{{\rm d}}
\def\e{{\rm e}}
\def\i{{\rm i}}

\Year{} %
\Month{} %
\Vol{} 
\No{} 
\BeginPage{1} 
\AuthorMark{{ }}  
\AuthorMarkCite{A. Author, B. Author, C. Author, and D. Author} 
\DOI{} 
\ArtNo{}
\title[Galactic Interstellar Scintillation observed from Four Globular
Cluster Pulsars by FAST]{Galactic Interstellar Scintillation Observed from Four Globular Cluster Pulsars by FAST}
\author[1,3]{Dandan Zhang}{}
\author[4,7]{Zhenzhao Tao}{}
\author[2]{Mao Yuan}{}
\author[5]{Jumei Yao}{}
\author[2,7]{Pei Wang}{}
\author[1,3]{Qijun Zhi}{qjzhi@gznu.edu.cn}
\author[2,7]{Weiwei Zhu}{zhuww@nao.cas.cn}
\author[6]{\\Xun Shi}{}
\author[8]{Michael Kramer}{}
\author[2,3,9,10]{Di Li}{}
\author[2]{Lei Zhang}{}
\author[6]{Guangxing Li}{}

\address[{\rm1}]{School of Mathematical Sciences, School of Physics and Electronic Sciences, Guizhou Normal University, Guiyang 550001, China;}
\address[{\rm2}]{National Astronomical Observatories, Chinese Academy of Sciences, Beijing 100012, China;}
\address[{\rm3}]{Guizhou Provincial Key Laboratory of Radio Astronomy and Data Processing, Guiyang 550001, China;}
\address[{\rm4}]{Department of Astronomy, Beijing Normal University, Beijing 100875, China;}
\address[{\rm5}]{Xinjiang Astronomical Observatories, Chinese Academy of Sciences, Xinjiang 830011, China;}
\address[{\rm6}]{South-Western Institute for Astronomy Research (SWIFAR), Yunnan University, Kunming 650000, China;}
\address[{\rm7}]{Institute for Frontiers in Astronomy and Astrophysics, Beijing Normal University,  Beijing 102206, China;}
\address[{\rm8}]{Max-Planck Institut für Radioastronomie, Auf dem Hügel 69, D-53121 Bonn, Germany;}
\address[{\rm9}]{Key Laboratory of Radio Astronomy and Technology, Chinese Academy of Sciences, Beijing 100101, China;}
\address[{\rm10}]{Computational Astronomy Platform, Zhejiang Lab, Hangzhou, Zhejiang 311121, China}

\maketitle \vspace{-3.5mm}{\footnotesize\begin{center} 
\end{center}}\vspace*{-5mm}
\begin{center}
\rule{16.5cm}{0.4pt}
\parbox{16.5cm}
{\begin{abstract}
We report detections of scintillation arcs for pulsars in globular clusters M5, M13 and M15 for the first time using the Five-hundred-meter Aperture Spherical radio Telescope (FAST). From observations of these arcs at multiple epochs, we infer that screen-like scattering medium exists at distances $4.1_{-0.3}^{+0.2}$~kpc, $6.7_{-0.2}^{+0.2}$~kpc and $1.3_{-1.0}^{+0.7}$~kpc from Earth in the directions of M5, M13 and M15, respectively. This means M5's and M13's scattering screens are located at $3.0_{-0.2}^{+0.1}$~kpc and $4.4_{-0.1}^{+0.1}$~kpc above the galactic plane, whereas, M15's is at $0.6_{-0.5}^{+0.3}$~kpc below the plane. We estimate the scintillation timescale and decorrelation bandwidth for each pulsar at each epoch using the one-dimensional auto-correlation in frequency and time of the dynamic spectra. We found that the boundary of the Local Bubble may have caused the scattering of M15, and detected the most distant off-plane scattering screens to date through pulsar scintillation, which provides evidence for understanding the medium circulation in the Milky Way.
\end{abstract}}
\end{center}\vspace*{-0.6cm}

\begin{center}
\parbox{16.5cm}
{\bf\jiuhao Pulsars, Scintillation, Globular clusters}
\end{center}

\begin{center}
{}
\Cit{~~~???, et al. ???. Sci China-Phys Mech Astron, 2014, 57: 1--6, doi:}
\end{center}

\textwidth=178truemm \textheight=236truemm

\wuhao\vspace*{1.5mm}

\begin{multicols}{2}

\renewcommand{\baselinestretch}{1.08} \baselineskip 12.2pt\parindent=10.8pt

\section{Introduction}\label{sec:1}
Pulsars scintillate because of a relative motion of the pulsar to the scintillation material in the interstellar medium (ISM) and the observer \cite{1990ARA&A..28..561R}. Pulsar scintillation observation is a unique tool for studying pulsars and the ISM. The pulsar scintillation dynamic spectrum, its intensity as a function of time and frequency, often shows crossed and periodic fringes \cite{1985MNRAS.213..167H,1986ApJ...307L..27C}. We can estimate the decorrelation bandwidth $\Delta v_{\rm d}$ and scintillation timescale $\Delta\ t_{\rm d}$ by using one-dimensional auto-correlation function (ACF) in frequency and time of the dynamic spectrum, and then verify different theories of turbulence in the interstellar medium \cite{2005MNRAS.358..270W}.

Stinebring et al. (2001)\cite{2001ApJ...549L..97S} first discovered the scintillation arc phenomenon in the secondary spectrum, the two-dimensional power spectrum of the dynamic spectrum. The secondary spectra occasionally exhibit parabolic arcs, which indicates the existence of thin-screen-like scattering medium. Subsequently, asymmetric, inverted and multiple parabolic arcs were detected from some pulsars\cite{2007ASPC..365..254S, 2010ApJ...708..232B, 2020RAA....20...76Y}. Through parabolic arcs, proper motion and distance of the pulsar, we can obtain the location of the scattering screens\cite{2020MNRAS.499.1468M, 2020ApJ...904..104R}.
For pulsars in binary systems, the scintillation arcs and the scintillation timescale may vary as the relative velocity of the pulsar changes due to orbital motion, thus, enable us to measure the orbital parameters \cite{1984Natur.310..300L, 2002ApJ...574L..75O, 2004ApJ...609L..71R, 2019MNRAS.485.4389R}. Scintillation observations are also valuable for understanding the influence of the interstellar medium on pulsar timing accuracy \cite{2019AAS...23314907B,2022MNRAS.511.1104M}. 

In this paper, owing to the sensitivity of FAST, we are able to detect scintillation arcs from pulsars in the globular clusters M5, M13 and M15 for the first time. These detections enabled us to localize a significant scattering medium between us and the globular clusters. We conducted interstellar scintillation analysis for FAST observations of PSRs J1518+0204A, J1518+0204C, J1641+3627A and J2129+1210A, which are located in M5, M13, and M15, respectively. We measured the distance to the scattering screen based on multiple measurements of the scintillation arc curvature, together with the distance and transverse velocity of each pulsar. The current arrangement of our work is as follows. In section 2, we introduce the observation and the data processing procedures. In section 3, we present the dynamic and secondary spectra, the scintillation parameters, the scintillation arcs, and the localization of the scattering screens. In section 4, we summarise our discussion and main conclusion.

\section{Observation and data reduction}  \label{sec:2}
The observations of globular clusters M5, M13, and M15 were carried out using the center beam of the 19-beam receiver of the FAST telescope. The FAST 19-beam receiver operates at a center frequency of 1250 MHz with a total bandwidth of 400 MHz and a system temperature of $\sim$20~K. The details of the 19-beam system are described by Jiang et al. (2020) \cite{2020RAA....20...64J}. We chose different frequency bands for each observation to minimize the contamination from radio frequency interference (RFI). Table \ref{tab:table1} displays the observation parameters of data, and the last column shows the frequency bands we used in this work.

We used DSPSR \cite{2011PASA...28....1V} to fold data and the \textit{pazi} command of the PSRCHIVE \cite{2012AR&T....9..237V} software package to remove RFI, and then used the \textit{psrflux} command to get the dynamic spectrum. We found that PSRs J1518+0204A, J1518+0204C, J1641+3627A, and J2129+1210A have more significant ‘fringes’ using a sub-integration time of 15~s for our pulsars.

\textbf{J1518+0204A} This is an isolated and strongest millisecond pulsar in M5, and its proper motion is $(\mu_{\alpha}, \mu_{\sigma})=(4.6 \pm 0.4, -8.9 \pm 1.0)$~mas~yr$^{-1}$ (the transverse velocity as ${V}_{\alpha}$=$163 \pm 32$~km~s$^{-1}$ and ${V}_{\sigma}$=$-316 \pm 36$~km~s$^{-1}$) as measured through timing \cite{2008ApJ...679.1433F}.

\textbf{J1518+0204C} This is a "Black Widow"(BW) system in M5 with an orbital period of 0.086829 days and a companion of minimum mass $\sim 0.04~ M_\odot$. Pallanca et al. (2014)\cite{2014ApJ...795...29P} measured the proper motion as $(\mu_{\alpha}, \mu_{\sigma})=(4.67 \pm 0.14, -8.24 \pm 0.36)$~mas~yr$^{-1}$ (the transverse velocity as ${V}_{\alpha}$=$166 \pm 12$~km~s$^{-1}$ and ${V}_{\sigma}$=$-292 \pm 13$~km~s$^{-1}$) through timing.

\textbf{J1641+3627A} This is the pulsar with the longest spin period of the six known pulsars in M13. Its transverse velocity ${V}_{\rm psr,\perp}$ is not currently known, however, the proper motion of the cluster M13 is $(\mu_{\alpha}, \mu_{\sigma})=(-3.18 \pm 0.01, -2.56 \pm 0.01)$~mas~yr$^{-1}$ (the transverse velocity as ${V}_{\alpha}$=$-112 \pm 2$~km~s$^{-1}$ and ${V}_{\sigma}$=$-90 \pm 1$~km~s$^{-1}$)\cite{2019MNRAS.482.5138B}. We will implicitly assume that the velocity of M13 is consistent with ${V}_{\rm psr,\perp}$ to calculate the distance of the screen.

\textbf{J2129+1210A} This is a pulsar very close to the core of M15. Kirsten et al. (2014) \cite{2014A&A...565A..43K} measured the proper motion as $(\mu_{\alpha}, \mu_{\sigma})=(0.54 \pm 0.14, -4.33 \pm 0.25)$~mas~yr$^{-1}$ (the transverse velocity as ${V}_{\alpha}$=$27 \pm 13$~km~s$^{-1}$ and ${V}_{\sigma}$=$-220 \pm 13$~km~s$^{-1}$) through very long baseline interferometry (VLBI).

Table \ref{tab:table2} shows parameters of each pulsar obtained from the ATNF (Australia Telescope National Facility)\footnote{\url{http://www.atnf.csiro.au/research/pulsar/psrcat/}} pulsar catalog. Column 3 shows the pulsar dispersion measure (DM) and column 4 shows the pulsar period. Column 5 shows the pulsar Orbital Period. Column 6 shows the distance of the globular cluster and column 7 shows the flux density. The last two columns are the transverse velocities of pulsars and globular clusters.

\begin{tablehere}
\footnotesize
\caption{Observation parameters of data}\label{tab:table1}
    \vspace{-1mm}\footnotesize
      \begin{center} \doublerulesep 0.1pt \tabcolsep 2pt
       \begin{tabular}[H]{lccccc}
        \hline
        Cluster & Date & Duration &Sample time & Channel width &Freq bands\\ 
        ~ & (MJD) & (min) & ($\mu \mathrm{s}$) &  (MHz) & (MHz)\\ 
        \hline
        \multirow{4}{*}{M5}   & 58828 &  240 & \multirow{4}{*}{49.152} & \multirow{4}{*}{ 0.122} & \multirow{4}{*}{1050-1150}\\
                    ~         & 58854 &  233 & ~ & ~ & ~ \\
                    ~         & 59113 &  118 &~ & ~ & ~ \\
                    ~         & 59626 &  120 & ~ & ~ & ~ \\
        \hdashline[3pt/3pt]            
        \multirow{4}{*}{M13}  & 58397 &  228 & \multirow{4}{*}{49.152}& \multirow{4}{*}{ 0.122} &\multirow{4}{*}{1050-1150}\\
                       ~      & 58685 &  60 & ~ & ~ &  ~ \\
                       ~      & 58723 &  60 &~ & ~ &  ~ \\
                       ~      & 58740 &  90 & ~ & ~ &  ~ \\
        \hdashline[3pt/3pt]
        \multirow{4}{*}{M15}  & 58831 &  122 & 196.608 & 0.061 & \multirow{6}{*}{1350-1450} \\
                       ~      & 59095 &  178 & 49.152 & 0.122 &  ~ \\
                        ~     & 59115 &  178 & 49.152 & 0.122 &  ~ \\
                       ~      & 59118 &  240 & 49.152 & 0.122 &  ~ \\
                       ~      & 59204 &  270 & 49.152 & 0.122 &  ~ \\
                       ~      & 59697 &  119 & 196.608 & 0.061 &  ~ \\
        \hline
		\end{tabular}
	  \end{center}
\end{tablehere}

\begin{table*}[!t]
\begin{threeparttable}
\caption{Pulsar parameters}\label{tab:table2}
\vspace{-1mm}\footnotesize
\begin{center} \doublerulesep 0.1pt \tabcolsep 10pt
\begin{tabular}{ccccccccc} 
\hline
 Cluster  &  PSR& DM & Period &Orbital Period & Dist\tnote{$^\dagger$}  & S1400 & ${V}_{\rm psr,\perp}$ & ${V}_{\rm Cluster,\perp}$\tnote{e} \\  
  ~      &  ~   & (cm$^{-3}$pc) &  (ms) &  (day) &  (kpc)   & (mJy) &  (km~s$^{-1}$) &  (km~s$^{-1}$)   \\  
\hline
\multirow{2}{*}{M5}  & J1518+0204A & 30.08 & 5.55 & \tnote{$\ast$}  & \multirow{2}{*}{$7.48 \pm 0.06$} & 0.12 &$355 \pm 32$\tnote{b}& \multirow{2}{*}{$379 \pm 3$}  \\

       ~            &  J1518+0204C & 29.31  & 2.48 & 0.087 & ~ & 0.039 & $336 \pm 12$\tnote{c}& \\
\hdashline[3pt/3pt]
M13& J1641+3627A  & 30.44 & 10.38 & \tnote{$\ast$}  & $7.42 \pm 0.08 $  & 0.14 &\tnote{$ \bullet $} & $144 \pm 2$ \\
M15  &J2129+1210A   & 67.31 &11.07 &\tnote{$\ast$}  &$10.71 \pm 0.10$& 0.2  &$222 \pm 13$ \tnote{d}& $196 \pm 2$\\
\hline
\end{tabular}
\begin{tablenotes}
\item[*] These pulsars are isolated
\item[{$^\dagger$}] These distances of globular clusters are obtained through a combination of Gaia Early Data Release 3 (EDR3) with distances based on Hubble Space
telescope (HST) data and literature based distances.\cite{2021MNRAS.505.5957B}
\item[b,c]These transverse velocities are measured through timing \cite{2008ApJ...679.1433F,2014ApJ...795...29P}.
\item[d]These transverse velocities are calculated through VLBI observations \cite{2014A&A...565A..43K}.
\item[$ \bullet $]No timing-based velocity has been published.
\item[e]These transverse velocities of globular clusters are measured via GAIA DR2 data\cite{2019MNRAS.482.5138B} .
\end{tablenotes}
\end{center}
\end{threeparttable}
\end{table*}

\section{Interstellar scintillation results}\label{sec:3}

In this section, we present the dynamic spectrum of each pulsar, and then compute the two-dimensional auto-covariance function (ACF) of the dynamic spectrum. In the end, we calculate the distance from the scattering screen to the globular cluster by fitting the scintillation arc in the secondary spectrum.

\subsection{Dynamic spectrum}\label{sec:3.1}

To obtain more pronounced dynamic spectra S(v,t), we folded each pulsar for small chunks of sub-integrations and subbands and used linear interpolation to fill the channels that were masked due to RFI. The dynamic spectrum was obtained by calculating the S/N of the pulsar in each sub-integration and subband. Supplementary Figure 1 shows the dynamic spectra of the four pulsars.

We folded PSRs J1518+0204A and J1518+0204C with a sub-integration time of 15~s and a channel bandwidth of 0.244~MHz. The dynamic spectrum of PSR J1518+0204A shows a distinct drift pattern in four observations due to refractive scintillation modulation\cite{1980MNRAS.192..799H}. The dynamic spectrum of PSR J1518+0204C shows regular eclipses about 15 minutes long at MJDs 58828, 58854 and 59113, and appears as criss-cross structures. We folded PSR J1641+3627A with a sub-integration time of 30~s and a channel bandwidth of 0.244~MHz. The scintillation timescale and decorrelation bandwidth of dynamic spectrum change dramatically over short time intervals, for example, from a narrow to a broad pattern in observations on MJDs 58723 and 58740. We folded PSR 2129+1210A with a sub-integration time of 15~s and a channel bandwidth of 0.122~MHz. The dynamic spectrum of this pulsar has narrow scintillation timescales and bandwidths.
\subsection{Auto-correlation function and scintillation parameters}\label{sec:3.2}
Following Cordes (1986) \cite{1986ApJ...311..183C} and Wang et al. (2008)  \cite{2008MNRAS.385.1393W}, we used the two-dimensional auto-correlation function (ACF) of the dynamic spectrum S(v,t) to measure the scintillation parameters. The ACF is defined as
\begin{equation}
C(\Delta v, \Delta t)=\sum_{v} \sum_{t} \Delta S(v, t) \Delta S(v+\Delta v, t+\Delta t),
\end{equation}
where  $\Delta S(v, t)=S(v, t)-\overline S$, $\overline S$ is the mean of the pulsar flux density, $v$ and $t$ are the frequency and the time, respectively. The normalized ACF is defined as
 \begin{equation}
\rho(\Delta v, \Delta t)=C(\Delta v, \Delta t)/C(0, 0),
\end{equation}
 and the decorrelation bandwidth $\Delta v_{\rm d}$ as the half-width at half-maximum of the ACF along the frequency lag axis, and the timescale $\Delta t_{\rm d}$ as the half-width at 1/e along the time lag axis\cite{1986ApJ...311..183C}.
Following Reardon et al. (2019)\cite{2019MNRAS.485.4389R}, we first perform a least square fit of the one-dimensional time-domain $C(0,\Delta t)$ to obtain $\Delta t_{\rm d}$. We fit $C(0,\Delta t)$ with
\begin{equation}
\label{E3}
C(0,\Delta t)=A~\rm exp~\left(-{\left|\frac{\Delta \textit{t}}{\Delta \textit{t}}_d \right|}^{\frac{5}{3}} \right),
\end{equation}
\begin{equation}
\label{E4}
C(0,0)=A+W,
\end{equation}
where $W$ is the white noise spike.
We keep $A$ and $W$ constant, and then fit the one-dimensional frequency-domain $C(\Delta v,0)$ with
\begin{equation}
\label{E5}
C(\Delta v,0)=A~\rm exp~\left(-{\left|\frac{\Delta \textit{v}}{\Delta \textit{v}_d/ln2} \right|} \right).
\end{equation}
The scintillation strength is defined as \cite{1990ARA&A..28..561R}
\begin{equation}
\label{E6}
u=\sqrt {f_{\rm obs}/\Delta \textit{v}_{\rm d}},
\end{equation}
where $f_{\rm obs}$ is the center frequency.
\begin{table*}[!t]
\caption{Analysis of dynamic spectrum and secondary spectra}\label{tab:table3}
\vspace{-1mm}\footnotesize
\begin{center} \doublerulesep 0.1pt \tabcolsep 10pt
       \begin{tabular}{cccccccccccc}
        \hline
        PSR & MJD  & $f_{\rm obs}$ &$A$ &$W$& $\Delta\ t_{\rm d}$ &$\Delta\ v_{\rm d}$ & $u$&S/N\\ 
        ~   & ~ &(MHz)&~ & ~&(min)        &  (MHz)     \\  
        \hline
    \multirow{4}{*}{J1518+0204A} &  58828&\multirow{4}{*}{1100} &$0.956 \pm 0.009$ &  $0.044 \pm 0.003$& $6.6 \pm 0.6$ & $ 2.8 \pm 0.3$  & $19.8 \pm 1.1$&10.9\\    
                             ~ &  58854& ~                    &$0.992 \pm 0.008$ &  $0.008 \pm 0.001$ & $5.8 \pm 0.6$ & $ 2.7 \pm 0.3$  & $20.2 \pm 1.1$&13.6\\
                             ~ &  59113& ~                    &$0.984 \pm 0.003$ &  $0.016 \pm 0.001$ & $4.0 \pm 0.5$ & $ 1.5 \pm 0.2$  & $27.1 \pm 1.8$&25.9\\
                             ~ &  59626& ~                    &$0.957 \pm 0.006$ &  $0.043 \pm 0.006$ & $6.5 \pm 0.7$ & $ 3.1 \pm 0.4$  & $18.8 \pm 1.2$&76.8\\
    \hdashline[3pt/3pt]                         
    \multirow{4}{*}{J1518+0204C} &  58828&\multirow{4}{*}{1100} &$0.980 \pm 0.007$ &  $0.020 \pm 0.002$ & $9.5 \pm 0.8$ & $ 3.3 \pm 0.4$  & $18.3 \pm 1.1$&8.2\\    
                             ~ &  58854& ~                    &$0.973 \pm 0.009$ &  $0.027 \pm 0.003$ & $8.0 \pm 0.6$ & $ 3.7 \pm 0.4$    & $17.2\pm 0.9$&16.9\\
                             ~ &  59113& ~                    &$0.987 \pm 0.006$ &  $0.013 \pm 0.001$ & $2.3 \pm 0.2$ & $ 1.0 \pm 0.1$   & $33.2 \pm 1.7$&10.9\\
                             ~ &  59626& ~                    &$0.905 \pm 0.005$ &  $0.095 \pm 0.002$ & $13.9 \pm 1.5$ & $ 6.7 \pm 0.7$  & $12.8 \pm 0.7$& 7.2\\ 
    \hdashline[3pt/3pt]                         
    \multirow{4}{*}{J1641+3627A} &  58397&\multirow{4}{*}{1100} &$0.949 \pm 0.004$ &  $0.051 \pm 0.002$ & $20.9 \pm 2.1$ & $ 4.8 \pm 0.5$   & $15.1 \pm 0.8$&8.2\\    
                             ~ &  58685& ~                    &$0.964 \pm 0.007$ &  $0.036 \pm 0.002$ & $10.3 \pm 1.4$ & $ 6.6 \pm 0.7$   & $12.9 \pm 0.7$&36.5 \\
                             ~ &  58723& ~                    &$0.961 \pm 0.006$ &  $0.039 \pm 0.002$ & $10.5 \pm 1.4$ & $ 3.0 \pm 0.4$   & $19.8 \pm 1.3$&15.6\\
                             ~ &  58740& ~                    &$0.983 \pm 0.005$ &  $0.017 \pm 0.002$ & $23.4 \pm 2.5$ & $ 10.2 \pm 1.2$  & $10.4 \pm 0.6$&65.5\\ 
    \hdashline[3pt/3pt]                         
                             
    \multirow{6}{*}{J2129+1210A}  ~ &  58831&\multirow{6}{*}{1400}  &$0.983 \pm 0.003$ &  $0.017 \pm 0.002$    & $2.6 \pm 0.4$ & $ 0.15 \pm 0.02$ & $93.5 \pm 5.8$&97.6\\
                             ~ &  59095& ~                  &$0.964 \pm 0.005$ &  $0.036 \pm 0.004$    & $2.0 \pm 0.3$ & $ 0.17 \pm 0.05$       & $81.6 \pm 7.8$ &85.2\\
                             ~ &  59115& ~                  &$0.961 \pm 0.008$ &  $0.039 \pm 0.005$    & $1.9 \pm 0.2$ &  $ 0.16 \pm 0.04 $      & $83.7 \pm 6.3$&90.8\\
                             ~ &  59118& ~                    &$0.985 \pm 0.006$ &  $0.015 \pm 0.001$    & $1.8 \pm 0.2$ & $ 0.13 \pm 0.04$      & $88.2 \pm 7.3$&88.2\\  
                             ~ &  59204& ~                    &$0.986 \pm 0.004$ &  $0.014 \pm 0.001$    & $3.4 \pm 0.5$ & $ 0.18 \pm 0.05$      &  $79.8 \pm 7.3$&94.0\\
                             ~ &  59697& ~                    &$0.983 \pm 0.007$ &  $0.017 \pm 0.002$    & $3.0 \pm 0.5$ & $ 0.24 \pm 0.04$     & $74.8 \pm 6.0$&91.8\\
    \hline
	\end{tabular}
	\end{center}
\begin{tablenotes}
\item[NOTE] Note. $A$, $W$, $\Delta\ t_{\rm d}$, $\Delta\ v_{\rm d}$ and u are obtained by 1D ACF. S/N is the  signal-to-noise ratio of the scintillation arc in secondary spectra.
\end{tablenotes}	
\end{table*}

For PSRs J1518+0204A and J1641+3627A (at MJDs 58685, 58723 and 58740), the 2D ACF of the dynamic spectrum shows some visible skewness attributed to phase gradients, i.e., a transverse gradient in the electron distribution somewhere along the line of sight, which varies but does not change direction between the observations. This skewness prevents us from directly estimating the decorrelation bandwidth $\Delta v_{\rm d}$ using Equations \ref{E5}. Therefore, we correct for the refractive shift according to the model described in Rickett et al. (2014) (Equation A6)\cite{2014ApJ...787..161R}. We then estimate the ACF parameters using  Equations \ref{E3} - \ref{E5}.
By contrast, the ACF skewness of the PSRs J1518+0204C, J1641+3627A (at MJD 58397) and J2129+1210A is much smaller.
We directly use Equations \ref{E3} - \ref{E5} to fit $A$, $W$, $\Delta t_{\rm d}$, and $\Delta v_{\rm d}$. The uncertainties of $\Delta t_{\rm d}$ and $\Delta v_{\rm d}$ include the uncertainty resulting from the fit and the `finite scintle error'\cite{ 2005MNRAS.358..270W}. Due to the overestimation of the decorrelation bandwidth $\Delta v_{\rm d}$ of PSR J2129+1210A with the 0.122 MHz frequency channels, we corrected it using Equation 3 of Wu et al. (2022)\cite{2022A&A...663A.116W}). Supplementary Figure 5 shows
the measured values of the decorrelation bandwidth of PSR
J2129+1210A and the corrected values in Table \ref{tab:table3}. We found that PSR J2129+1210A has narrow scintillation timescales ($\sim2.5 \pm 0.4$~min) and bandwidths ($\sim 0.17 \pm 0.04$~MHz). Supplementary Figure 2 - 5 shows the 2D ACFs before (panel a) and after (panel a1) correction, the 1D ACFs in frequency (panel b and b1 ) and time (panel c). Table \ref{tab:table3} provides the ACF parameters. The derived scintillation strength, $u$, indicates that the scattering is strong along the path to each globular cluster. 

\subsection{Secondary spectra} \label{sec:3.3}

The secondary spectrum $P(f_{\rm v},f_{\rm t})$ is the two-dimensional Fourier transform of the dynamic spectrum\cite{2004MNRAS.354...43W,2006ApJ...637..346C}. Two scattered waves from directions $\theta_1$ and $\theta_2$ interfering with each other produce a fringe pattern. The conjugate frequency ($f_{\rm v}$) and conjugate time ($f_{\rm t}$) in the secondary spectrum are
\begin{equation}
\label{E7}
f_{\rm v}=\left[\frac{D(1-s)}{2c s}\right]\left(\boldsymbol{\theta}_{2}^{2}-\boldsymbol{\theta}_{1}^{2}\right),
\end{equation}
\begin{equation}
\label{E8}
f_{\rm t}=\left(\frac{1}{\lambda s}\right)\left(\boldsymbol{\theta}_{2}-\boldsymbol{\theta}_{1}\right) \cdot \boldsymbol{V}_{\rm eff,\perp},
\end{equation}
where $s=D_{\rm p}/D$, $D_{\rm p}$ is the distance from the pulsar to the
scattering screen, $\lambda$ is the observing wavelength and $c$ is the speed of light, $D$ is the distance from the observer to the pulsar, $\boldsymbol{V}_{\rm eff,\perp}$ is an effective velocity of the velocities of the pulsar and observer relative to the screen,
\begin{equation}
\label{E9}
\boldsymbol{V}_{\rm eff,\perp}=(1-s)\boldsymbol{V}_{\rm psr,\perp}+s\boldsymbol{V}_{\rm obs,\perp}-\boldsymbol{V}_{\rm screen,\perp},
\end{equation}
and contains only the transverse components of each velocity \cite{1998ApJ...507..846C}.
Following Stinebring et al. (2001)\cite{2001ApJ...549L..97S}, the parabolic arc in the secondary spectrum is
\begin{equation}
\label{E10}
f_{\rm v}=\eta f_{\rm t}^{2},
\end{equation}
where $\eta$ is the parabolic curvature 
\begin{equation}
\label{E11}
\eta=4625 \frac{D_{\mathrm{kpc}}{s}({1-s})}{ f_{\rm obs}^{2} (\lvert\boldsymbol{V}_{\mathrm{\rm eff,} \perp}\rvert \rm cos\phi_{s}) ^{2}},
\end{equation}
where $f_{\rm obs}$ is the observing frequency in GHz, and $\phi_{\rm s}$ is the angle between the major axis of image and the effective velocity vector when the scattering screen is anisotropy \cite{2014ApJ...787..161R,2005AAS...20719504R}. We assume the globular cluster distance as the pulsar distance to calculate the scattering screen distance. Following this, we can derive the distance of the scattering screen to the globular cluster to be $Ds$ in kpc. 

Before measuring the arc curvature $\eta$, we present a quantitative analysis of the signal-to-noise ratio (S/N) of the scintillation arc in Table \ref{tab:table3}. We define S/N as the ratio of the scintillation variance to the noise variance\cite{2022ApJ...941...34S}. The scintillation region is visually selected to be significantly above the noise floor. For PSRs J1518+0204A, J1518+0204C and J1641+3627A, the scintillation region as $\vert f_{\rm t} \vert <4$, the noise region as $7<f_{\rm t} <15$. For PSR J2129+1210A, scintillation region as $\vert f_{\rm t} \vert<6$. The noise region as $ 8<f_{\rm t} <20$.   

\begin{figure}[H]
\centering
\includegraphics[width=3.3in]{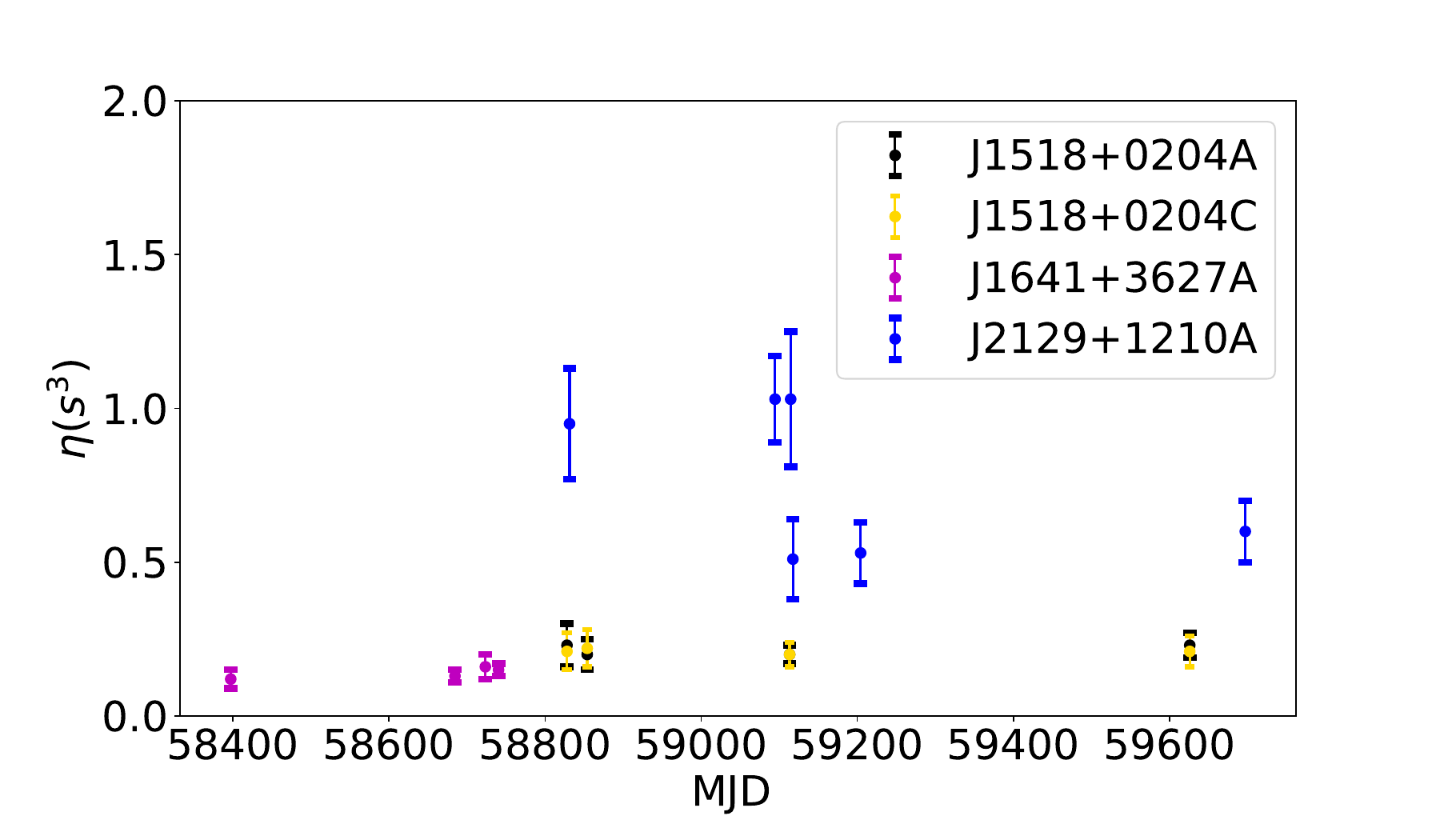}
\caption{The arc curvature $\eta$ as a function of time. The arc curvatures of PSRs J1518+0204A and J1518+0204C are consistent.}
\label{fig:Figure1}
\end{figure}

To estimate the arc curvature measurement $\eta$, following Reardon et al. (2020) \cite{2020ApJ...904..104R}, we re-sample the secondary spectrum to obtain a "normalized" secondary spectrum, and then average along $f_{\rm v}$ to obtain the Doppler profile. We re-scale the x-axis of the Doppler profile into physical units of the $\eta$, and then detect peaks in mean power. The left panel of Supplementary Figure 6 - 9 shows the mean power as a function of arc curvature. We smooth the mean power using the  savgol\_filter from scipy \footnote{\url{https://docs.scipy.org/doc/scipy/reference/generated/scipy.signal.savgol\_filter.html}} to find a reasonable fitting peak region. We set the length of the filter window to be 0.7 $\%$ of the number of steps in etas through several experimental tests. The uncertainty of $\eta$ is determined from the noise level in the secondary spectrum . The left panel of Supplementary Figure 6 - 9 shows the results obtained for the arc curvature $\eta$. For PSRs J1518+0204A, J1518+0204C and J1641+3627A, the arc curvature $\eta$ does not change between epochs. However, the arc curvature $\eta$ of PSR J2129+1210A varies at each epoch (Figure\ref{fig:Figure1}), suggesting a scattering screen near Earth.

\begin{figure*}[!t]
\centering
\subfigure[J1518+0204A]{
\includegraphics[width=7.5cm]{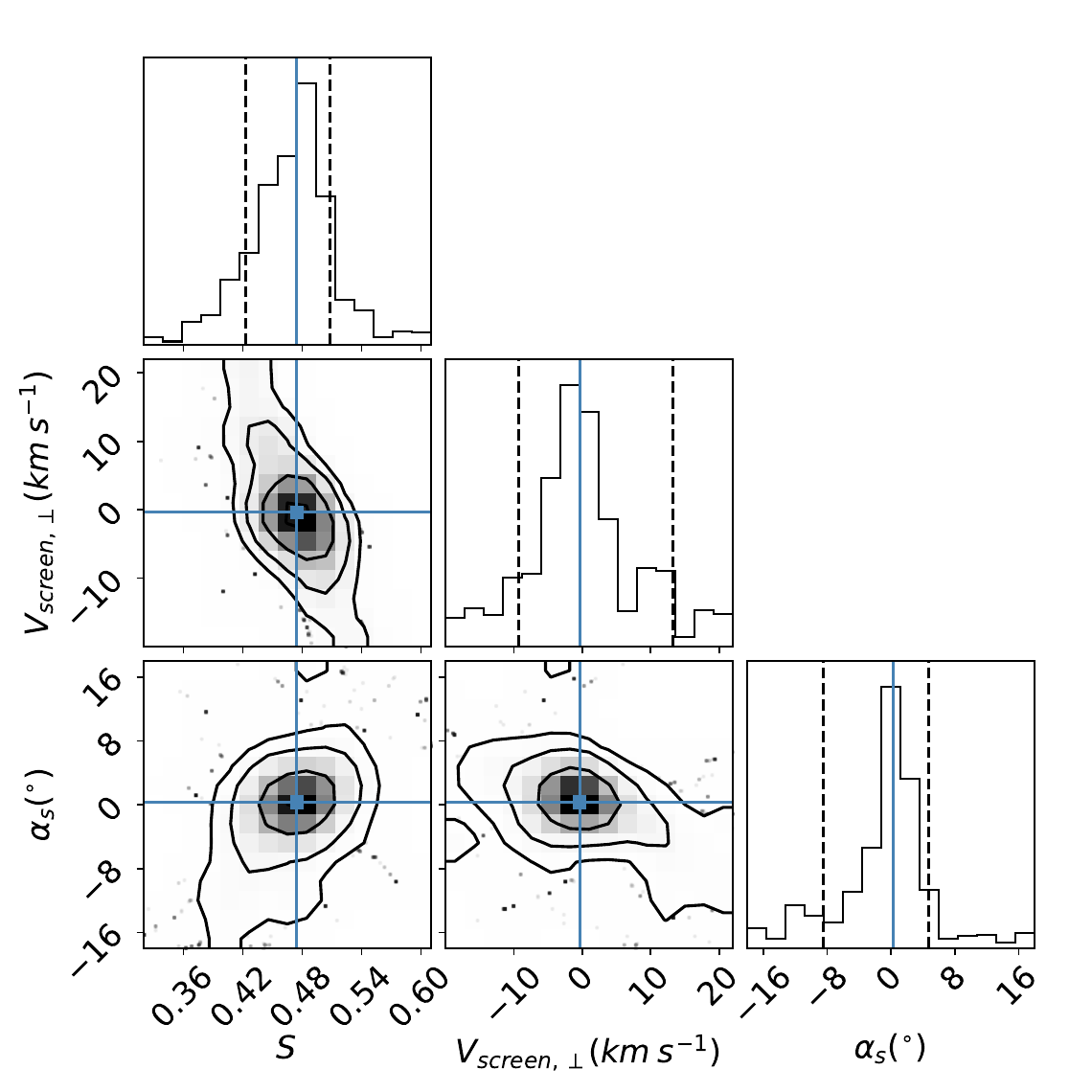}
}
\qquad
\subfigure[J1518+0204C]{
\includegraphics[width=7.5cm]{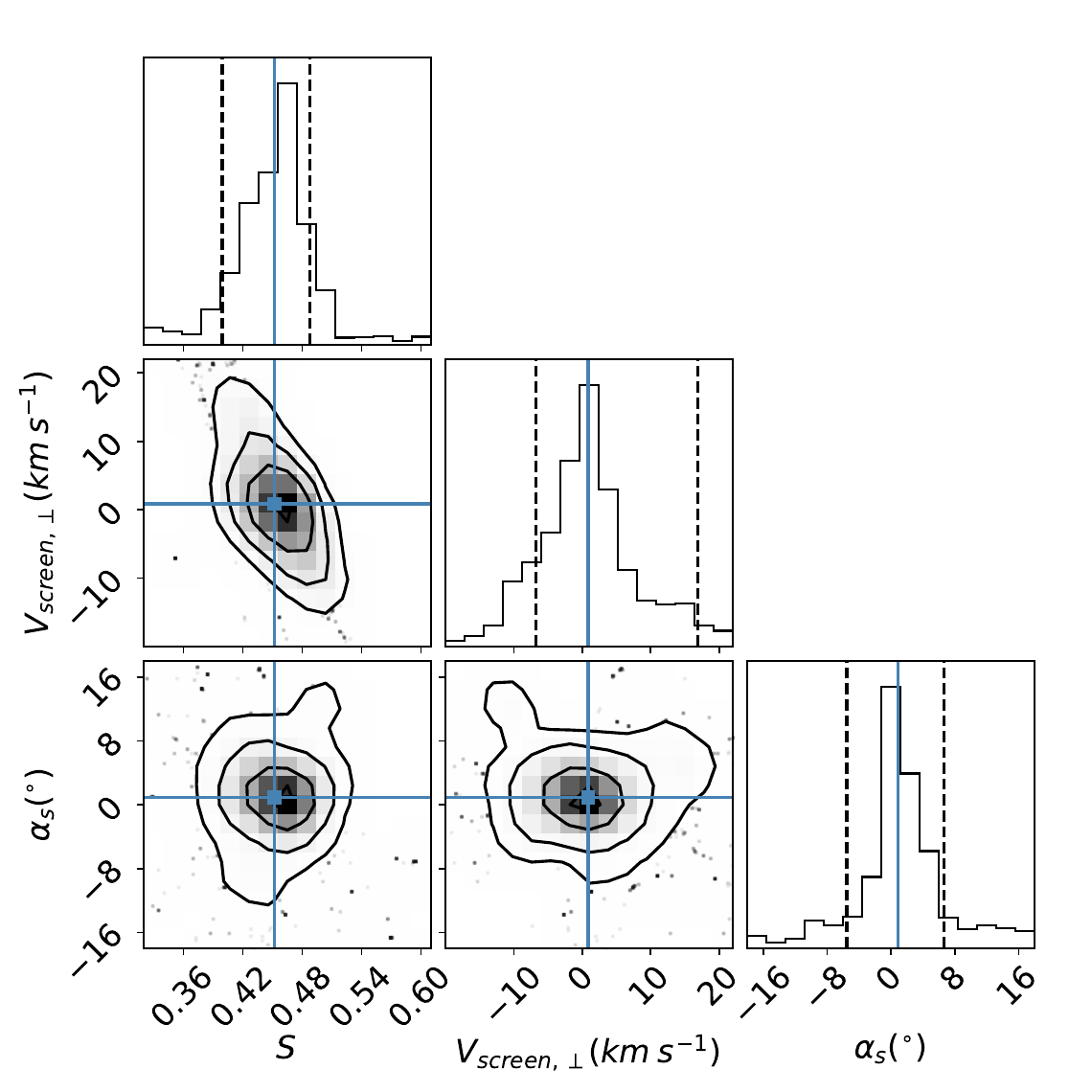}
}
\qquad
\subfigure[J1641+3627A]{
\includegraphics[width=7.5cm]{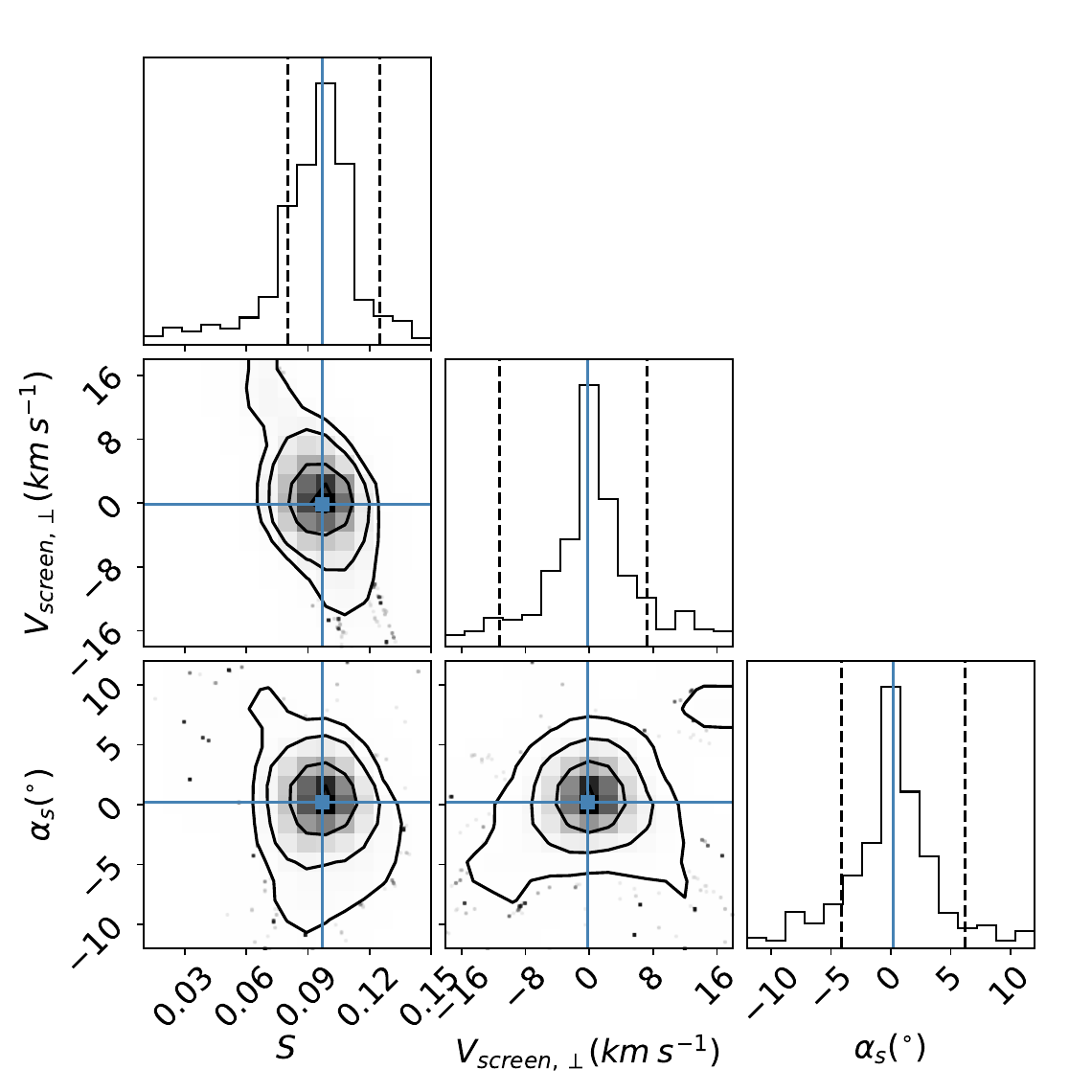}
}
\qquad
\subfigure[J2129+1210A]{
\includegraphics[width=7.5cm]{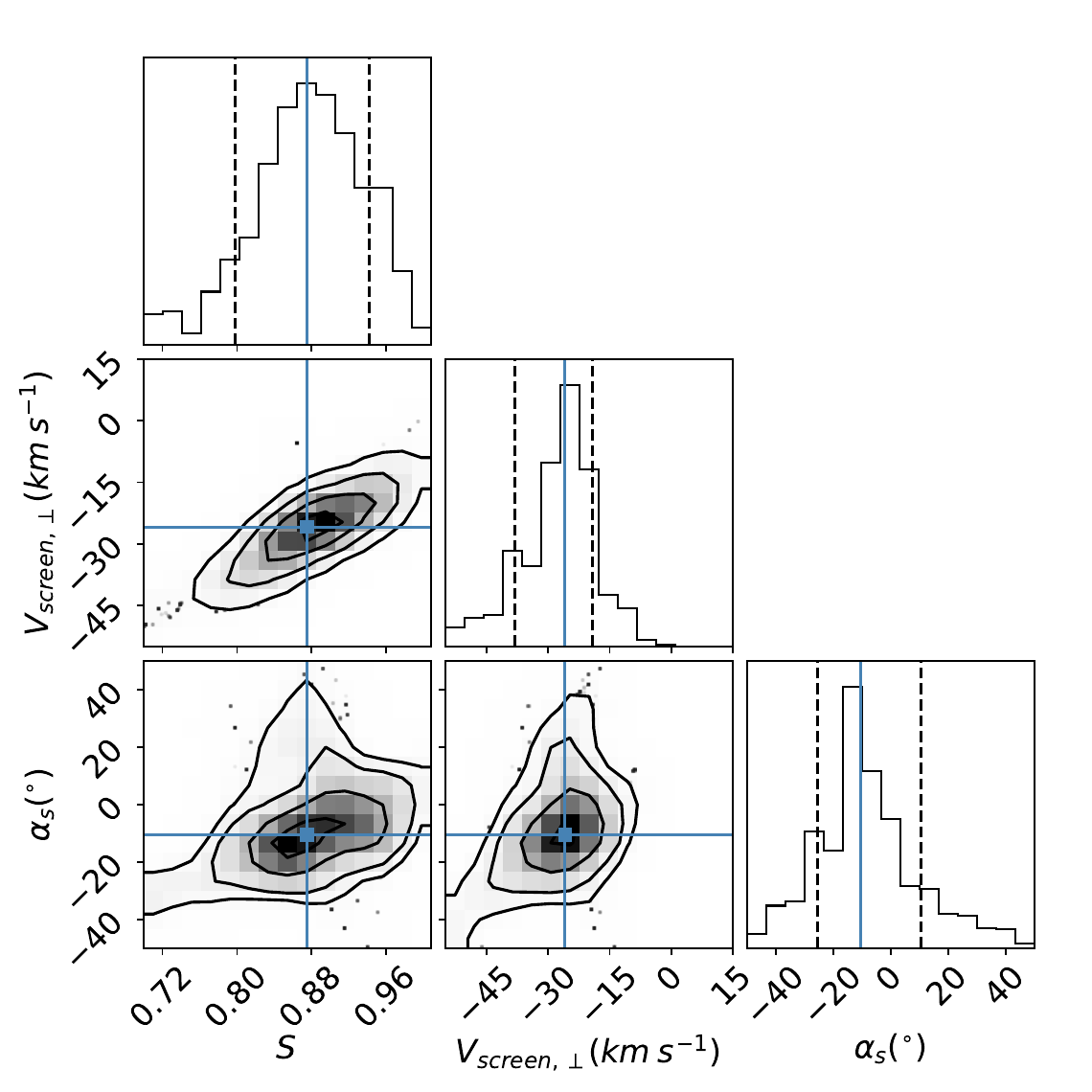}
}
\caption{Corner plots of MCMC results of four pulsars ((a) J1518+0204A, (b) J1518+0204C, (c) J1641+3627A, (d) J2129+1210A) show the parameter space for the scattering screen model. We used 5000 iterations and adopted a uniform distribution for the priors distribution of the model parameters.
Vertical dashed lines indicate the 0.16, 0.84 quantiles. Blue solid lines represent the distribution mean. Multiple observations of the scintillation arc curvature allow us to constrain all screen parameters, namely the screen velocity $V_{\rm screen,\perp}$, the angle {$\alpha_{\rm s}$}, and the fractional distance distance to the scattering screen $s$.}
\label{fig:Figure2}
\end{figure*}

Following McKee et al. (2022)\cite{2022ApJ...927...99M}, we assume that the scattering screen is highly anisotropic, and the effective velocity is parallel to the major axis of the scattered images. Therefore, the $\boldsymbol{V}_{\mathrm{screen,} \perp }$ in Equation \ref{E9} becomes a scalar along the major axis of the image. We fit the arc curvature measurement $\eta$ using the following functional:
\begin{equation}
\label{E12}
\begin{split}
&\eta(t,s,V_{\rm screen, \perp},\alpha_{\rm s}) \\
&=4625 \frac{D_{\mathrm{kpc}}{s}({1-s})}{ f_{\mathrm{GHz}}^{2} (\lvert \boldsymbol{A}\rvert \rm cos(\alpha_{\rm s}-\alpha_{\rm v})-V_{\rm {screen,}\perp }) ^{2}},
\end{split}
\end{equation}
where 
\begin{equation}
\label{E13}
\boldsymbol{A}=(1-s)\boldsymbol{V}_{\mathrm{psr,} \perp }+s\boldsymbol{V}_{\mathrm{obs,} \perp}(t)
,
\end{equation}
$\alpha_{\rm v}$ is the angle between $\boldsymbol{A}$ and the decl. axis, $\alpha_{\rm s}$ is the angle between the major axis of the image and the decl. axis, so $\alpha_{\rm s}$-$\alpha_{\rm v}$ is the angle between the $\boldsymbol{A}$ and the major axis of the image,
$\boldsymbol{V}_{\mathrm{psr,} \perp }$, $\boldsymbol{V}_{\mathrm{obs,} \perp }$ and ${V}_{\mathrm{screen,} \perp }$ is in $\rm km~s^{-1}$, $\eta$ is in $s^3$, 

We estimated the parameters of the model using a Markov chain Monte Carlo (MCMC) technique\cite{2022ApJ...927...99M}. The velocity of each pulsar is provided in Section \ref{sec:2}, and we calculated the transverse components of Earth's velocity with respect to each pulsar using the CALCEPH packages\footnote{\url{ https://juliapackages.com/p/calceph/}}. The $\alpha_{\rm v}$ can be calculated from Equation \ref{E13}.
We obtained the unknown scattering screen parameters: $s$, $V_{\rm screen,\perp}$ and $\alpha_{\rm s}$. The corner plots of MCMC results are shown in Figure \ref{fig:Figure2}. Table \ref{tab:table4} listed the scattering screen parameters that we obtained from fitting the model using Equation \ref{E12} and Equation \ref{E13}.

\begin{figure*}[!t]
\centering
\includegraphics[width=5.0in]{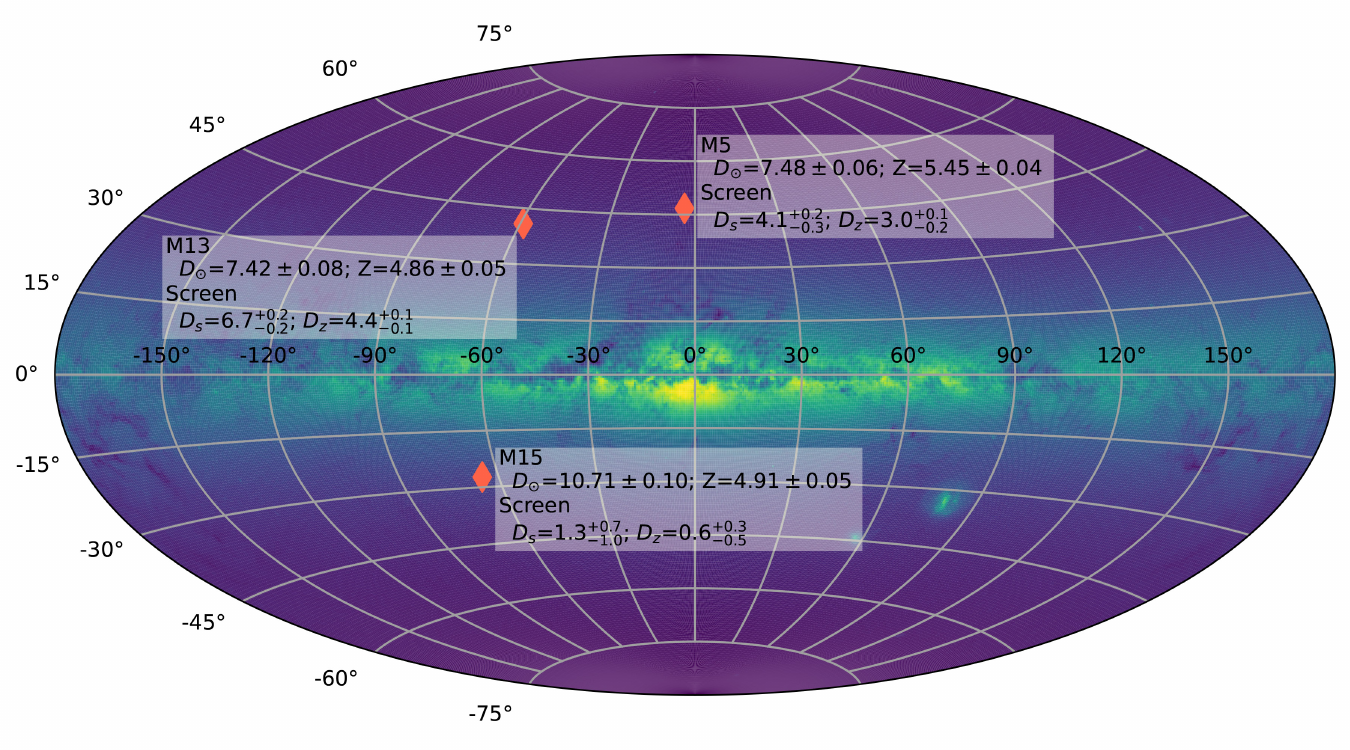}
\caption{Distribution of M5, M13 and M15 in the Milky Way. $D_{\odot}$ is the distance from the globular clusters to the Earth. $Z$ is the height from the globular clusters to the plane. $D_{s}$ is the distance from the scattering screens to the Earth. $D_z$ is the height from the scattering screens to the plane. The unit of all values is kpc.}

\label{fig:Figure3}
\end{figure*}

The secondary spectra of PSRs J1518+0204A and J1518+0204C display clear parabolic arcs with deep unfilled valleys, indicating that their scattering screens are anisotropic\cite{2022ApJ...941...34S}. The constrained screen parameters for these two pulsars in M5 (PSRs J1518+0204A and J1518+0204C) are consistent with each other. This indicates that the parabolic arcs of these two pulsars are likely caused by a common scattering screen with an extent of greater than 40.5 arc seconds ($ 4.1_{-0.3}^{+0.2}$~kpc at a fractional screen distance of $s=0.46_{-0.03}^{+0.04}$). For M13 and M15, their scattering screens are located at $6.7_{-0.2}^{+0.2}$~kpc and $1.3_{-1.0}^{+0.7}$~kpc away from the Earth, respectively. We show the positions of the globular clusters and the scattering screens within the Milky Way in Figure \ref{fig:Figure3}\footnote{The background picture is modified from an images by ESA/Gaia/DPAC.}. We find that M5's and M13's scattering screens are located at $ 3.0_{-0.2}^{+0.1}$~kpc, $ 4.4_{-0.1}^{+0.1}$~kpc above the galactic plane, and M15's scattering screens is located at $ 0.6_{-0.5}^{+0.3}$~kpc below the galactic plane.
\begin{tablehere}
\begin{threeparttable}
\renewcommand\arraystretch{1.6}
\caption{Fitted scattering screen parameters are obtained by fitting the MCMC model.}\label{tab:table4}
\vspace{-1mm}\footnotesize
\begin{center} \doublerulesep 0.1pt \tabcolsep 4pt
\begin{tabular}{ccccccccc} 
\hline
  PSR& s &$\alpha_{\rm s}~(^{\circ})$ &$V_{\rm screen, \perp}~(\rm km~s^{-1})$ &$D_{\rm s}~(\rm kpc)$\tnote{1} &$D_{\rm z}~(\rm kpc)$\tnote{2}\\  
\hline
 
 J1518+0204A &$0.47_{-0.05}^{+0.03}$ & $0.3_{-9}^{+4}$ &  $-0.4_{-9}^{+14}$  & $ 4.0_{-0.4}^{+0.2}$& $ 2.9_{-0.3}^{+0.1}$\\
 J1518+0204C & $0.45_{-0.05}^{+0.04}$ & $0.9_{-6}^{+6}$ &  $0.9_{-8}^{+16}$ &  $ 4.1_{-0.4}^{+0.3}$&  $ 3.0_{-0.3}^{+0.2}$\\
 J1641+3627A &$0.10_{-0.02}^{+0.03}$ & $0.2_{-5}^{+4}$ &  $-0.2_{-10}^{+9}$  &$6.7_{-0.2}^{+0.2}$&  $ 4.4_{-0.1}^{+0.1}$ \\
 J2129+1210A & $0.88_{-0.09}^{+0.07}$ & $-11_{-15}^{+21}$ &  $-26_{-12}^{+7}$  &  $1.3_{-1.0}^{+0.7}$& $ 0.6_{-0.5}^{+0.3}$ \\

\hline
\end{tabular}
\begin{tablenotes}
\item[1]The distance from the scattering screens to the Earth.
\item[2]The height from the scattering screens to the galactic plane 
\end{tablenotes}
\end{center}
\end{threeparttable}
\end{tablehere}

\section{Discussion and conclusions}\label{sec:3}
Using FAST observation, we detected interstellar scintillation for PSR J1518+0204A, PSR J1518+0204C, PSR J1641+3627A and PSR J2129+1210A located in the globular clusters M5, M13 and M15, respectively. We found the dynamic spectra to vary in different observations, along with the scintillation timescale $\Delta t_{\rm d}$ and decorrelation bandwidth $\Delta v_{\rm d}$. The quantitative analysis of these variations is given in Table \ref{tab:table3}.
It is not uncommon for scintillation parameters to exhibit variation over time\cite{2022A&A...664A.116L,2016ApJ...818..166L}.
The scintillation timescale $\Delta t_{\rm d}$ and decorrelation bandwidth $\Delta v_{\rm d}$ depend on the scale of the diffractive scintillation pattern, $l_{\rm d}$\cite{2022A&A...664A.116L},  
\begin{equation}
\label{E14}
\Delta t_{\rm d}= \frac{l_{\rm d}}{V_\mathrm{ISS}},
\end{equation}
\begin{equation}
\label{E15}
\Delta V_{\rm d} \propto l_{\rm d}^2f_{\rm obs}^2,
\end{equation}
where ${V_\mathrm{ISS}}={V}_{\rm eff,\perp}/s$ for a thin screen. Thus, a large $s$ indicates that the scattering screen is close to Earth, and the transverse velocity of Earth causes variations in $\Delta t_{\rm d}$. Equations \ref{E14} and \ref{E15} can also be used to explain the variations of both $\Delta t_{\rm d}$ and $\Delta v_{\rm d}$ when the $l_{\rm d}$ changes unpredictably.

Through scintillation arc curvatures, we can locate the positions of the dense interstellar medium in the Galaxy and reveal a part of the galactic structure. In general, the closer the scattering screen is to the Earth, the greater the variations of arc curvatures. 
According to a recent census of scattering screens\cite{2022A&A...663A.116W, 2022ApJ...941...34S, 2023MNRAS.518.1086M}, before this work, the most distant scattering screen from Earth is from PSR J1141-6545 \cite{2019MNRAS.485.4389R} with a distance of $7_{-3}^{+4}$~kpc, and the furthest off-plane scattering screen is from PSR J1543-0620 with a Galactic height of $0.9 \pm 0.2$~kpc \cite{2022ApJ...941...34S}.
As reported in this work, the scattering screens in front of M5 and M13 pulsars are the furthest off-plane screens discovered so far. Their distances above the galactic plane ($3.0_{-0.2}^{+0.1}$~kpc and $ 4.4_{-0.1}^{+0.1}$~kpc). At such heights, it is nearly impossible for these screens to be associated with H~II regions or individual supernova remnants, i.e. the commonly suggested scattering source of some pulsars\cite{1976ApJ...205..762S}. The scale height of the thick disk of the warm interstellar medium is estimated to be only around 1.7~kpc\cite{2017ApJ...835...29Y}. 
Thus these screens are likely \textsl{not} associated with some small-scale genetic structures of the warm interstellar medium, such as the reconnection sheets as suggested by Pen $\&$ Levin (2014)\cite{2014MNRAS.442.3338P}. 
One interesting possibility is that the screens are related to the global matter circulation of our Galaxy. 
It has been proposed that clusters of supernova explosions can break through the galactic disk opening up a tube-like chimney, injecting hot gas into the galactic halo and driving a “galactic fountain”. After that, the gas cools, condenses, and eventually falls back onto the galactic disk like rainfall \cite{2019ApJ...887...89W,2020ApJ...898..148L}. The vertical boundary of the chimney and the interface between the high-latitude cold gas and the galactic halo are viable candidates for these observed scattering screens.

For the globular clusters M15, we find that the arc curvature $\eta$ of PSR J2129+1210A varies considerably between epochs. We obtain that the M15's scattering screen is located at $1.3_{-1.0}^{+0.7}$~kpc from Earth. The Earth is located in the Local Bubble with a diameter of approximately 0.3~kpc\cite{2022Natur.601..334Z}. So, it is possible that M15's scattering screen arises from the boundary of the Local Bubble.

\vspace*{2mm} \Acknowledgements{\bahao 
This work was supported by the National SKA Program of China (Grant
Nos. 2020SKA0120200, 2022SKA0130100, and 2022SKA0130104), the
National Nature Science Foundation of China (Grant Nos. 12273008,
11873067, 12041303, 12041304, 61875087, U1831120, U1838106,
61803373, 11303069, 11373011, 11873080, U2031117, and 12103069), the
Natural Science and Technology Foundation of Guizhou Province (Grant
No. [2023]024), the National Key R$\&$D Program of China (Grant No.
2017YFB0503300), the Guizhou Provincial Science and Technology Foundation (Grant Nos. ZK[2023]024, ZK[[2022]304, [2017]5726-37, and
[2018]5769-02), the Major Science and Technology Program of Xinjiang
Uygur Autonomous Region (Grant Nos. 2022A03013-4, and 2022A03013-
2), the Scientific Research Project of the Guizhou Provincial Education
(Grant Nos. KY[2022]132, and KY[2022]137), the Guizhou Province Science and Technology Support Program (General Project) (Grant No. Qianhe
Support [2023] General 333), the Foundation of Guizhou Provincial Education Department (Grant No. KY (2020) 003), the Natural Science Foundation of Xinjiiang Uygur Autonomous Region (Grant No. 2022D01D85),
the Youth Innovation Promotion Association CAS (Grant No. 2021055), the
CAS Project for Young Scientists in Basic Research (Grant No. YSBR-006),
the Cultivation Project for FAST Scientific Payoff and Research Achievement
of CAMS-CAS and ACAMAR Postdoctoral Fellowship. This work made use
of data from the FAST. FAST is a Chinese national mega-science facility,
built and operated by the National Astronomical Observatories, Chinese
Academy of Sciences. The authors also thank Lin Wang, Zhichen Pan and Pengfei Wang for proposing and acquiring the FAST data.}

\end{multicols}
\end{document}